\documentclass[%
aps,
prx,
reprint,
superscriptaddress,
nofootinbib,
amsmath,
amssymb
]{revtex4-2}
\usepackage[T1]{fontenc}
\usepackage[utf8]{inputenc} 
\usepackage{
    phfqit,
    graphicx,
    multirow,
    tabularx,
    placeins, 
    nicefrac, 
    hyperref, 
    threeparttable, 
    enumitem,
    makecell,
    comment,
}

\usepackage[print-unity-mantissa=false]{siunitx}
\usepackage{mathtools}\mathtoolsset{centercolon}
\usepackage{fixmath}
\usepackage{phfparen}
\usepackage{mathcommand}
\usepackage{stmaryrd}
\usepackage{orcidlink}
\LoopCommands{\lettersAll{CG}{DCS}{MSPE}{NLL}{SE}}{\declaremathcommand#2{\textnormal{#1}}}
\LoopCommands{
{pr}{wt}
}{\DeclareMathOperator#2{#1}}

\usepackage{algorithm}
\usepackage{algpseudocodex}
\usepackage{amsthm}
\usepackage[capitalize]{cleveref}
\crefname{section}{Section}{Sections}
\Crefname{section}{Section}{Sections}

\crefname{app}{Appendix}{Appendices}
\Crefname{app}{Appendix}{Appendices}

\theoremstyle{definition}

\theoremstyle{remark}

\usepackage[dvipsnames]{xcolor}
\hypersetup{
    colorlinks,
    linkcolor={blue!50!black},
    citecolor={blue!50!black},
    urlcolor={blue!80!black}
}

\usepackage[caption=false]{subfig} 
\DeclarePairedDelimiterX\pbraket[2]{\langle\!\langle}{\rangle\!\rangle}{#1 \delimsize\vert #2}

\begin{document}

\title{A route to damage tolerance exceeding $10\%$ \\ in shuttling-equipped quantum processors}

\newcommand{\qmaddress}{\affiliation{Quantum Motion, 9 Sterling Way, London N7 9HJ, United Kingdom}}
\newcommand{\oxaddress}{\affiliation{Department of Materials, University of Oxford, Parks Road, Oxford OX1 3PH, United Kingdom}}
\newcommand{\imecaddress}{\affiliation{Imec, Kapeldreef 75, 3001 Leuven, Belgium}}
\newcommand{\kuladdress}{\affiliation{Department of Electrical Engineering, KU Leuven, Kasteelpark Arenberg 10, 3001 Leuven, Belgium}}
\newcommand{\ghentaddress}{\affiliation{Department of Physics and Astronomy, Ghent University, Krijgslaan 281, 9000 Gent, Belgium}}

\author{Quinten Eggerickx\,\orcidlink{0000-0002-4709-3115}}
\email{quinten.eggerickx@imec.be}
\qmaddress
\imecaddress
\kuladdress
\ghentaddress
\author{Simon C. Benjamin\,\orcidlink{0000-0002-7766-5348}}
\email{simon.benjamin@materials.ox.ac.uk}
\qmaddress
\oxaddress

\date{\today}

\begin{abstract}
This is a short study of an approach offering high tolerance to damage (i.e. defects or `drop outs') in solid state fault-tolerant quantum computing. Our method is primarily aimed at semiconductor electron spin-qubit systems, which have been shown to support fast and high-fidelity shuttling along pre-defined paths. We adapt the recent CAbLECAR method of Chadwick and Chong: stabilisers are performed by ancillas which each follow a bespoke pre-programmed path. We consider the simple surface code but we damage the physical lattice, and rely on route-solving software to find efficient pathways under constraints enforcing stabiliser commutation and hook error avoidance. Solutions are then converted to detector error models for Stim and logical error rates are obtained. We express our results by gauging the logical performance against that of a
pristine lattice, using the notion of a reduced \emph{equivalent} surface-code
distance; for reasonable underlying error rates we find that $10\%$ damage leaves roughly half of the pristine equivalent distance ($d_\text{equiv}\approx0.48\,d_\text{pristine}$ in the large-array limit, rising to $\approx0.60$ for our smallest array). This suggests that one can tolerate substantial damage by building oversized arrays. We note that investigating damage tolerance of other qLDPC codes is a straightforward generalisation, and potentially one could adapt to damage emerging at runtime.
\end{abstract}

\maketitle

\section{Introduction} \label{sec:intro}

In this paper we explore the question of how well a fault-tolerant solid-state quantum computer can tolerate damage, by which we mean physically inoperable components. This has been well-explored for systems where qubits are fixed in place, but is more of an open question for systems where qubits have shuttling -- the ability to move, but not without some finite cost in time and noise. We are especially interested in what we call latticework architectures, which are 2D networks of linear segments along which qubits can move. Such networks are highly relevant to semiconductor quantum dot spin qubit systems. We restrict our attention to the surface code in memory mode, but the approach has broad applicability and we remark on generalisations in \cref{sec:future}. 

Our approach follows in the spirit of a recent paper~\cite{cablecar2026} and involves assigning to each mobile qubit a specific movement schedule that can be unique. Equivalently we can say that each elementary shuttling segment will flow either one way or the other, or be static, on each time step so that qubits realise their schedules. These behaviours are determined before runtime and programmed in, so that at runtime the sequence is followed systematically. Given a unique damaged latticework and a scheduling solution, we derive a corresponding detector error model for Stim and evaluate the logical performance of that system. We gather statistics over different latticework sizes and severities of damage; with certain reasonable choices for noise (and importantly, for shuttling noise) we find that $10\%$ damage reduces the equivalent code distance to between about
$60\%$ (smallest array) and $48\%$ (largest array) of the pristine value.
Crucially this fraction converges as the array grows, rather than falling without
bound, so the reduction can be compensated by over-sizing the original device by
a roughly constant factor --- about twofold linear increase at $10\%$ damage. 

\section{Background} \label{sec:background}

Fault tolerant quantum computing (FTQC) involves quantum error correcting codes (QECCs) which are monitored by repeated stabiliser measurements, with the information being processed by decoders. The code's design is typically orientated primarily for resilience to noise that occurs at runtime due to environmental interactions and imperfect qubit control or measurement. However, in many systems there may also be damage to the hardware -- sites where qubits, single- or two-qubit gates, measurements, etc. are simply inoperable or deemed too noisy to contribute to operations.  Such damage, variously called `hard defects' or `dropouts', might be known and characterised before runtime, or it might arise from issues (e.g. cosmic ray impact) that occur mid-execution and persist for multiple stabiliser rounds. 

Various studies have explored the means to tolerate finite levels of damage, and have characterised the eventual cost in terms of increased logical error risk for a given size of system. Substantial effort has gone into methods that vary the code itself to effectively quarantine the damaged region~\cite{stace2009thresholds,nagayama2017defective,auger2017fabrication,strikis2023scalable,siegel2023adaptive,leroux2025snakes,debroy2025luci},
for example by applying new stabiliser measurements around its
perimeter. This method has the important advantage that it can solve the issue with in-place qubits: no new wiring or physical relocation of qubits is needed provided that the underlying hardware has appropriate topology. 

One relevant example is the `Snakes and Ladders' approach of Leroux et al~\cite{leroux2025snakes}, where the authors explored the surface code and applied damage to three key types of components (data qubits, ancilla qubits, couplers). The authors studied the way that a logical qubit's protection against physical errors falls with increasing damage, and characterised this with an effective code distance that can be expressed as a fraction of the distance that {\em would be achieved} with a damage-free system. The authors found that for $1\%$ damage to all components, the effective distance settles to around $80\%$ of the ideal as logical size increases. This is good news in the sense that a designer could simply oversize the logical qubits in the machine, anticipating a reduction of this kind. However, as the damage increases much beyond $1\%$ the effective distance drops precipitously and good solutions for given logical patches may be impossible to find. Other approaches may reduce the impact through more complex stabiliser sequences~\cite{debroy2025luci}, and meanwhile defect tolerant surface-code operation has been demonstrated on superconducting
hardware~\cite{drouet2026error}. However, tolerating damage levels well above $1\%$ seems very challenging with such approaches.

In contrast to systems such as superconducting qubits, where the physical systems are fixed in place, some hardware modalities involve physical qubits that can shuttle. Neutral atoms and ion traps of course exploit this fully, but electron spins in silicon quantum dot devices also support shuttling. Recently hardware demonstrations have shown excellent speed and fidelity, with negligible risk of qubit loss. Shuttling is along linear paths, but such paths can meet at junctions so as to cover a 2D surface with a latticework. 

Early demonstrations transported spins dot-to-dot in `bucket brigade' mode:
coherent transfer of an electron spin through a quantum-dot array in
isotopically enriched silicon with $99.4\%$ average coherent-transfer
fidelity~\cite{Yoneda2021}, a two-qubit gate between distant spins enacted by
shuttling~\cite{Noiri2022shuttle}, and repeated shuttling through extended dot
arrays~\cite{Zwerver2023}. More recently `conveyor-mode' shuttling, in which a
smoothly moving potential minimum carries the electron
continuously~\cite{Langrock2023}, has emerged as the scalable variant:
spin-coherent conveyor transport has been demonstrated in
Si/SiGe~\cite{Struck2024} and in germanium~\cite{van2024coherent}, and
has culminated in the transport of a single spin over $10\,\mu$m in under
$200\,$ns ($\sim\!50$\,m/s) at $\sim\!99.5\%$ fidelity~\cite{de2025high}.
Conveyor shuttling through a T-junction --- the elementary requirement for the two-dimensional latticeworks we consider --- has also been demonstrated experimentally~\cite{beer2026conveyor}, while pulse-shaping theory predicts per-hop error rates could fall to the $10^{-5}$ level~\cite{oda2026suppressing}.

A number of papers have proposed architectures that directly rely on shuttling
either in one dimension~\cite{kunne2024spinbus,Siegel2024,trilinear2025,moncy2026surface} or in
two~\cite{Vandersypen2017,Boter2022,cablecar2026}. Typically authors have
initially targetted the surface code because of its well-understood stabiliser
requirements~\cite{Siegel2024,moncy2026surface}, although alternatives have been
noted~\cite{cablecar2026,moncy2026surface} and some architectures have been shown to
support radically changed or extended surface code
entities~\cite{siegel2025snakes,cai2023looped}.

A recent paper~\cite{cablecar2026} from Chadwick and Chong introduced the CAbLECAR paradigm and explored how a general-purpose regular shuttling latticework, designed without targetting any specific code, can be programmed to support any given qLDPC code by designating the motion of each ancilla step-by-step. The authors report that codes with superior rates can in fact outperform the surface code on such a latticework, provided that the shuttling noise is low enough (as expected given that the more complex codes generally need longer shuttling routes). Interestingly the authors rely on a route-finding algorithm to automate the considerable challenge of finding efficient schedules for the ancillas, which must of course avoid collisions. 

The approach taken by the CAbLECAR paper is adopted and modified in the present paper. Instead of contrasting the performance of a variety of codes on a given latticework, we use the same kind of techniques to explore the way the surface code can operate on a damaged latticework. We limit our study to the case that damage is a static, pre-determined characteristic of the hardware -- thus the problem of finding a legitimate schedule for the ancilla movement can be solved before runtime. While we follow the spirit of the CAbLECAR paper, our reference latticework and its rules, together with our route solving and evaluation tools, were written from scratch for this project. 

In \cref{sec:model} we specify the simple model that we use in our study, and in \cref{sec:routing} we briefly explain our route solving approach and its priorities. In \cref{sec:stim} we explain the mapping from a mooted solution to the Stim model that can test its logical performance including our simple error model. Our results (\cref{sec:results}) presents the performance that we observe, and its sensitivity to the error rate assumptions. Finally in \cref{sec:discussion} we discuss the generalisations including the application to a range of codes (as in the CAbLECAR paper) to explore per-code damage robustness, as well as the latticework variations and relaxing the demand for per-path directional control. 

\section{The latticework and damage model}
\label{sec:model}

\begin{figure*}
  \centering
  \includegraphics[width=\textwidth]{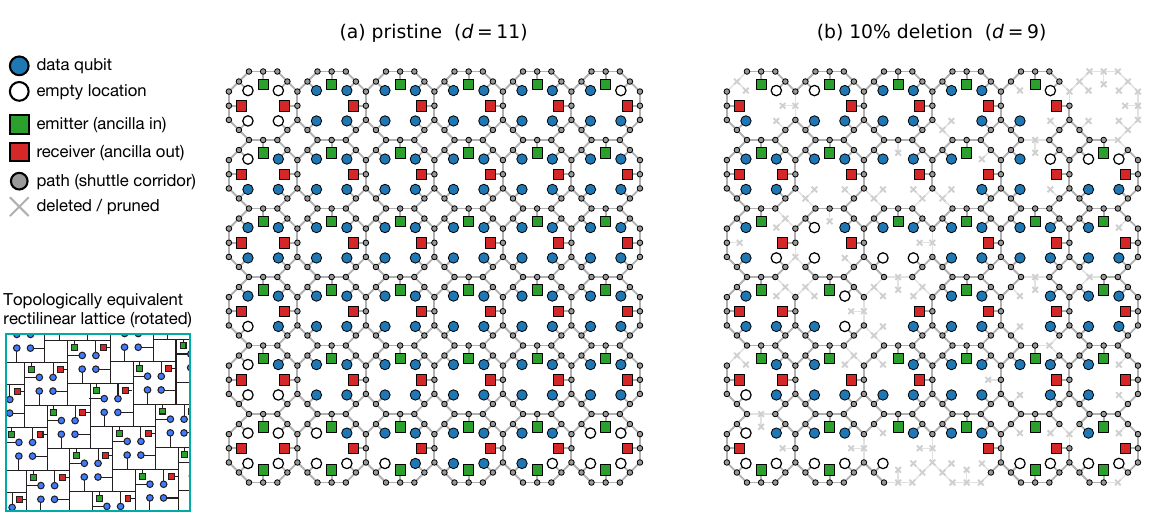}
  \caption{The truncated-square (octagon) latticework, pristine~(a) and
  after $10\%$ node deletion~(b). Filled blue circles are \emph{location}
  nodes occupied by a data qubit; white circles are unoccupied locations.
  Green and red squares are \emph{emitter} and \emph{receiver} nodes, where
  ancillas are injected and read out; small grey circles are \emph{path}
  nodes forming the shuttle corridors. Deleted nodes, and nodes pruned
  because deletion disconnected them from the main component, are shown as
  light crosses. The pristine $6\times6$ latticework hosts the full
  distance-$11$ code ($121$ data qubits); on the damaged instance shown,
  distance~$9$ is the largest code that fits ($81$ data qubits). The inset
  (lower left) redraws the pristine latticework as a topologically equivalent
  rectilinear lattice (shown rotated), a geometry that may be a more natural
  fabrication target.}
  \label{fig:lattice}
\end{figure*}

Our reference latticework is a $G\times G$ array of octagonal cells in a
truncated-square tiling (\cref{fig:lattice}(a)). Each cell contributes four
kinds of node. The four corner-cut edges of the octagon carry \emph{location}
nodes: sites where a data qubit may reside and onto which an ancilla can step
to interact with it. The flat north/south edges carry
\emph{emitter} nodes, at which fresh ancillas are injected, and the flat
east/west edges carry \emph{receiver} nodes, at which ancillas are measured
and removed. Finally, the octagon vertices and edge midpoints are \emph{path}
nodes: the shuttle corridors along which ancillas move. Location, emitter and
receiver nodes are always leaves hanging off a path node, so all transport
happens on the path network. A $G\times G$ array provides a $2G\times 2G$
grid of location sites and can therefore host a rotated surface code of
distance up to $d_{\max}=2G-1$.

\begin{figure}
  \centering
  \includegraphics[width=\columnwidth]{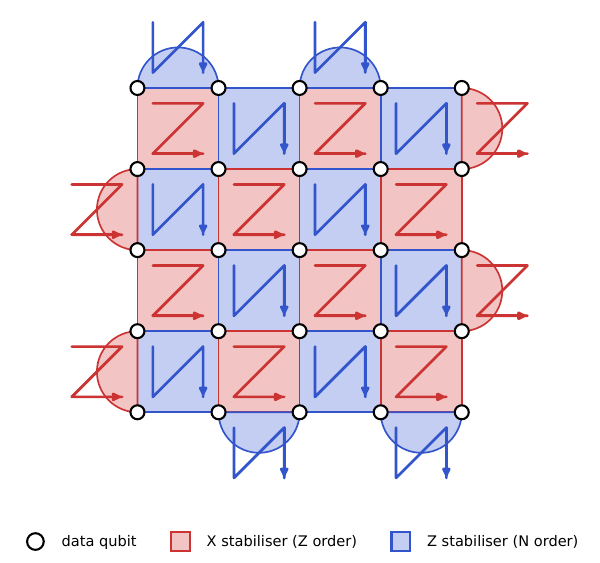}
  \caption{The distance-$d$ rotated surface code (shown for $d=5$): $d^2$
  data qubits, with $d^2-1$ X (red) and Z (blue) stabilisers in a
  checkerboard, and weight-two boundary stabilisers as lobes. The
  hook-error-avoiding N--Z syndrome-extraction
  ordering~\cite{tomita2014low,heim2016optimal} is shown for each stabilizer: X-type ancillas visit their corner data
  qubits in the ``Z''-shaped order NW$\to$NE$\to$SW$\to$SE, Z-type ancillas
  in the ``N''-shaped order NW$\to$SW$\to$NE$\to$SE. The four steps form one
  global schedule; boundary stabilisers simply skip their missing corners.}
  \label{fig:code}
\end{figure}

\paragraph{Damage.} We model hard defects by deleting a fraction $f$ of the
lattice nodes --- exactly $\lfloor fN \rceil$ of the $N$ nodes, drawn
uniformly at random over \emph{all} node types --- and then retaining only
the largest connected component, since fragments cut off from the main
network can play no part in the computation. Damage is static and known
before runtime, so all placement and routing decisions below are made
offline. Each damaged instance is identified by $(G, f, \mathrm{seed})$ and
is exactly reproducible. In the representative $f=0.1$ instance of \cref{fig:lattice}(b), $62$ of the $624$ nodes are deleted directly and a further $42$ are pruned by the connectivity requirement.

\paragraph{Code placement.} To host a distance-$d$ code
(\cref{fig:code}) the $d^2$ data qubits are mapped onto surviving
location sites by a connectivity-aware placement whose aim is to keep the
schedule length a property of the code distance rather than of the lattice
size. Because the placement is driven by connectivity rather than geometry,
neighbouring code qubits need not occupy geometrically adjacent sites on a
damaged lattice, as visible in Fig.~\ref{fig:lattice}(b). A distance $d$
is \emph{structurally infeasible} on an instance if fewer than $d^2$ location
sites survive, or no valid placement exists.

\section{Route solving}
\label{sec:routing}

One round of syndrome extraction requires each of the $d^{2}-1$ stabiliser ancillas to execute an \emph{itinerary}: injection at an emitter node, a shuttled tour that steps onto and off each of its two or four data-qubit location nodes in turn, and read-out at a receiver node. Data qubits are visited in the
hook-error-avoiding N--Z order of \cref{fig:code}. Since the N--Z
ordering is a single global four-step schedule, it also fixes the order in
which the several ancillas sharing a given data qubit must interact with it;
these orderings are imposed on the router as precedence constraints.

Routes are found on the time-expanded lattice graph: time advances in
discrete steps, every edge traversal (hop) takes one step, and capacity
constraints forbid two ancillas from occupying the same node or edge
simultaneously. We use a greedy solver: ancillas are routed sequentially,
each by a space--time breadth-first search that finds its
earliest-completion itinerary while treating all previously routed ancillas
as moving obstacles and respecting the precedence constraints. The solver
returns each ancilla's full trajectory (its position at every time step) and
read-out time; the duration of the complete round for all ancillas is the \emph{makespan}
$M$. Greedy sequential routing is not optimal, but it is fast enough to
evaluate the thousands of damaged instances required below, and the same
solver is used for damaged and pristine lattices alike, so comparisons are
like for like.

One timestep is the duration of transiting a single shuttling edge; gate,
initialisation and measurement operations are taken to be instantaneous on this
scale. This is less of an idealisation than it appears. A stabiliser
interaction requires the ancilla to step from its path node onto the data
qubit's location node and back, so each gate already occupies two shuttling
timesteps that the makespan counts; only the two-qubit operation itself is
treated as instantaneous. Initialisation and measurement occur at the emitter
and receiver nodes and can be overlapped in time with shuttling elsewhere on the latticework --- with their throughput raised simply by adding more emitters and receivers --- so they need not lengthen the round. The idling exposure that governs our error budget is therefore set by the hop-count makespan.

If the solver cannot complete a round, the distance is \emph{routing
infeasible} on that instance. Damage therefore acts on the code twice: it removes location sites, shrinking the largest structurally fitting distance, and it reduces the density of valid paths. For the distances that do still fit, this second effect lengthens ancilla tours and the makespan, and so increases the noise accumulated per round.

\section{Mapping to Stim, and the error model}
\label{sec:stim}

\paragraph{Circuit construction.} A routed instance is compiled into a
$d$-round memory-experiment circuit in Stim~\cite{Gidney2021stim}, with the
same one-round routing repeated in every round. The compilation follows the
schedule literally: gates occur at the time steps at which the router placed
the corresponding ancilla--data interactions, and the ancilla trajectories
determine, per qubit, every hop and every idle interval.

\paragraph{Error model.}

All gate, initialization, and measurement errors are held at a standard
circuit-level value $p = 10^{-3}$. Concretely, each operation is followed by
the corresponding noise channel:
\begin{itemize}
  \item \textbf{Initialization} ($R$): an $X$ error with probability $p$
        ($X_{\mathrm{ERROR}}(p)$).
  \item \textbf{Single-qubit gates}: a single-qubit depolarizing channel
        $\mathrm{DEPOLARIZE1}(p)$.
  \item \textbf{Two-qubit gates}: a two-qubit depolarizing channel
        $\mathrm{DEPOLARIZE2}(p)$.
  \item \textbf{Measurement} ($M$): the reported outcome is flipped with
        probability $p$.
  \item \textbf{Shuttle edge transit
  }: pure dephasing on the shuttled qubit,
        $Z_{\mathrm{ERROR}}(p_{\mathrm{shuttle}})$.
  \item \textbf{Idling}: for each timestep in which a qubit genuinely waits
        --- neither shuttling nor participating in a gate --- pure dephasing
        $Z_{\mathrm{ERROR}}\big((1-e^{-\Delta t/T_2})/2\big)$, with $\Delta t$ one
        timestep and $T_2$ the dephasing time.
\end{itemize}
The gate, initialization, and measurement strengths are specified in the following text; they coincide with the typical magnitude used in surface-code models, and are of the order of the single- and two-qubit
fidelities already demonstrated in silicon spin qubits (which have surpassed
the $99\%$ threshold across several implementations)~\cite{takeda2020resonantly,mills2022two,weinstein2023universal,wu2025simultaneous,broz2026demonstration,steinacker2025industry,mkadzik2025operating}, so that our results
isolate the effect of the shuttling latticework rather than of any unusually
optimistic 1Q or 2Q operation set.

The two channels specific to a shuttling architecture --- the shuttle and
idling errors --- are both modelled as \emph{pure dephasing}, the dominant
decoherence mechanism for spin qubits transported through a quantum-dot array.
We parameterize the idling channel by the error accrued over an ideal
$M_{\mathrm{ideal}}=32$-timestep makespan, $p_{\mathrm{idle}}$, setting
$T_2 = -M_{\mathrm{ideal}}/\ln(1-2p_{\mathrm{idle}})$; our baseline is
$p_{\mathrm{idle}}=10^{-3}$. Because the logical-$X$ observable is directly
sensitive to $Z$ errors, this makes the simulation conservatively responsive
to exactly the noise that shuttling introduces.

Both shuttling parameters can be grounded in demonstrated hardware.
High-fidelity single-spin shuttling in silicon has reached $\sim99.5\%$
fidelity while transporting a spin over $10\,\mu$m in under $200\,$ns, i.e.\ at
$\sim50\,\mathrm{m/s}$~\cite{de2025high}; theoretical analysis predicts that
pulse shaping could push shuttling errors to the $\sim10^{-5}$ level (with
speeds up to hundreds of m/s)~\cite{oda2026suppressing}. Our primary value
$p_{\mathrm{shuttle}}=2\times10^{-4}$ (with $10^{-4}$ reported as a prospective
improved-shuttling case) therefore brackets current and near-term performance.
The idling parameter follows similarly: at $\sim50\,\mathrm{m/s}$ a
$1$--$2\,\mu$m hop takes $20$--$40\,$ns, so the ideal $32$-timestep makespan
spans $\tau\sim0.6$--$1.3\,\mu$s. Silicon spin-qubit dephasing times reach tens
of milliseconds under dynamical decoupling \cite{veldhorst2014addressable}; taking a representative $T_2\sim10\,$ms, the idling
error over one ideal makespan is $(1-e^{-\tau/T_2})/2\approx5\times10^{-5}$. Our
baseline $p_{\mathrm{idle}}=10^{-3}$ is thus roughly 20 times more
pessimistic, so the idling assumptions are conservative.

\paragraph{Logical error rate.} Circuits are sampled and decoded with
minimum-weight perfect matching (PyMatching~\cite{Higgott2022pymatching}) on
the circuit's detector error model. Because idling is pure dephasing, the
logical-$X$ ($\ket{+}$-memory) error rate dominates the logical-$Z$ rate for
every parameter set studied here, and we report the logical-$X$ error rate
per $d$-round experiment throughout. Shots are drawn in batches until a target number of logical failures is observed, subject to a per-evaluation shot cap. The target is $100$ for deterministic single-instance evaluations and $10$ for damaged ensemble samples, whose precision is instead controlled at the ensemble level. With $h$ observed failures, the relative standard error is $\simeq 1/\sqrt{h}$.

\paragraph{Best distance per instance.} On a damaged lattice a larger code
is not necessarily better: a distance that barely fits routes through a
constricted network, and its extra shuttling and idling noise can outweigh
the added code distance. For each instance we therefore evaluate
\emph{every} odd distance $3\le d\le 2G-1$ that fits, and take the instance's performance to be
$\mathrm{LER}^{*}=\min_d \mathrm{LER}(d)$, attained at the best distance
$d^{*}$.

\paragraph{Ensemble sampling.} For each deletion fraction $f>0$ we draw independent damaged instances until the relative standard error of the mean (SEM) of $\mathrm{LER}^*$ falls below $10\%$ (at least $20$, at most $300$ instances). The $f=0$ point is deterministic and is evaluated
once at the higher precision.

\paragraph{Equivalent distance.} To express residual protection in intuitive
units we map each instance's $\mathrm{LER}^{*}$ to an \emph{equivalent
distance}: the distance at which a pristine code, under identical noise,
reaches the same logical error rate. The reference curve evaluates the
pristine code at every odd $d'\le 2G-1$, each on its minimal latticework
$G'=(d'+1)/2$; well below threshold $\ln\mathrm{LER}$ is linear in $d$, so
we fit $\ln\mathrm{LER}=a+b\,d'$ (typically $R^2\gtrsim0.97$) and invert,
$d_{\mathrm{eq}}=(\ln\mathrm{LER}^{*}-a)/b$. The reference is rebuilt for
every noise-parameter set, so damaged and pristine codes are always compared
under identical noise. We use ``equivalent distance'' deliberately: it is a
fixed-noise comparison against pristine codes, related to but distinct from
the effective-distance measure of Ref.~\cite{leroux2025snakes}. Near
threshold the log-linear fit degrades and the mapping loses meaning; such
points are reported only qualitatively.


\section{Results}
\label{sec:results}

\subsection{Choice of operating point}
\label{sec:operating-point}

Before presenting the damage results we assign the two shuttling-specific noise
parameters. Within the range of physically plausible noise we select values sufficiently strong that even the smallest array's LER is appreciably affected by these errors. In this way, the impact of these forms of noise with the increasing demands of shuttling on larger damaged arrays will be clearly exposed.

\begin{figure}[t]
  \centering
  \includegraphics[width=\linewidth]{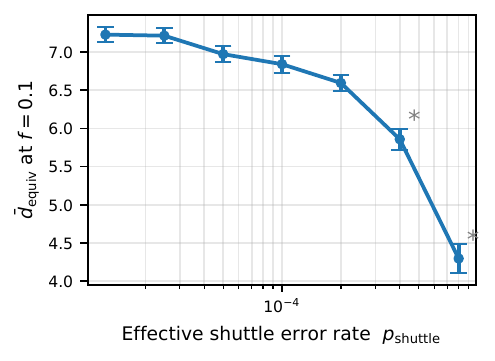}
  \caption{Mean equivalent distance $\bar d_{\rm equiv}$ at $10\%$ deletion
  ($6\times6$ grid, $X$ basis) as a function of the dephasing shuttle error rate.
  Each point maps the damaged samples' best LER onto an equivalent pristine
  distance via the same-noise pristine fit; error bars are the standard error of
  the mean (SEM). The curve is nearly flat below $\sim\!2\times10^{-4}$ and falls
  steeply above it, where the pristine log-linear fit also begins to break down
  ($R^2$ dropping from $0.99$ to $0.86$) as the code approaches threshold.
  Asterisks mark the operating points where this equivalent-distance mapping is
  becoming unreliable --- a degraded pristine fit ($R^2<0.95$) and/or some
  samples whose LER falls beyond the pristine reference range and must be
  extrapolated.}

  \label{fig:shuttle-effect}
\end{figure}

\cref{fig:shuttle-effect} sweeps the dephasing shuttle rate at a fixed
$10\%$ deletion fraction on the 6$\times$6 grid. The mean equivalent distance is
essentially flat across the low-rate range --- $7.23$ at $1.25\times10^{-5}$
down to $6.59$ at $2\times10^{-4}$, a loss of under $0.7$ in equivalent distance
over more than a decade of shuttling error rate. Beyond $2\times10^{-4}$ it falls sharply
($5.86$ at $4\times10^{-4}$, $4.30$ at $8\times10^{-4}$), and the pristine fit
quality degrades in the same window, signalling that the equivalent-distance
mapping itself is losing meaning as the code nears threshold. The knee therefore
sits at $p_{\mathrm{shuttle}}\approx2\times10^{-4}$, and we adopt this as our
primary value: it is the largest shuttle error still on the plateau, so the
results are neither optimistic nor past the cliff. The prospective
improved-shuttling value $10^{-4}$ (\cref{sec:improved-shuttling}) lies just
inside the plateau.

\begin{figure}[t]
  \centering
  \includegraphics[width=\linewidth]{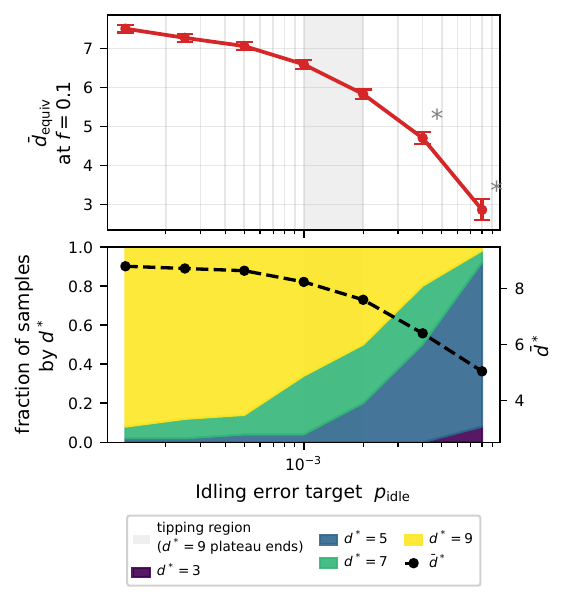}
  \caption{Idling sensitivity at $10\%$ deletion ($6\times6$ grid), with the
  shuttle rate fixed at $2\times10^{-4}$. \emph{Top:} mean equivalent distance
  $\bar{d}_{\text{equiv}}$ vs the idling target $p_{\mathrm{idle}}$ (error bars
  are the SEM). \emph{Bottom:} the composition of the winning code size $d^*$
  across samples, with $\bar{d}^*$ overlaid. The shaded band marks the tipping
  region ($1\times10^{-3}$--$2\times10^{-3}$) where the winning distance departs
  its low-noise $d=9$ plateau and the equivalent distance tips down. Asterisks
  (top panel) mark points where the equivalent-distance mapping is becoming
  unreliable --- a degraded pristine fit ($R^2<0.95$) and/or extrapolated
  samples --- as the code nears threshold, exactly as in
  Fig.~\ref{fig:shuttle-effect}.}

  \label{fig:idle-effect}
\end{figure}

\cref{fig:idle-effect} fixes the shuttle rate at this primary value and
sweeps the idling target. Below $\sim10^{-3}$ the winning code size sits on a
stable $d=9$ plateau; between $10^{-3}$ and $2\times10^{-3}$ the winning distance
collapses off that plateau and the equivalent distance tips downward in lockstep.
We adopt $p_{\mathrm{idle}}=10^{-3}$ as our baseline --- the low edge of this
tipping region, i.e.\ the largest idling error still short of the
collapse. As established in the error model, this baseline is already about 20 times more pessimistic than the $\sim 5 \times 10^{-5}$ implied by
demonstrated silicon $T_2$ times, so it is a conservative operating point in
absolute terms as well.

\subsection{Damage across grid sizes}
\label{sec:main-result}

\begin{figure*}[t]
  \centering
  \includegraphics[width=\linewidth]{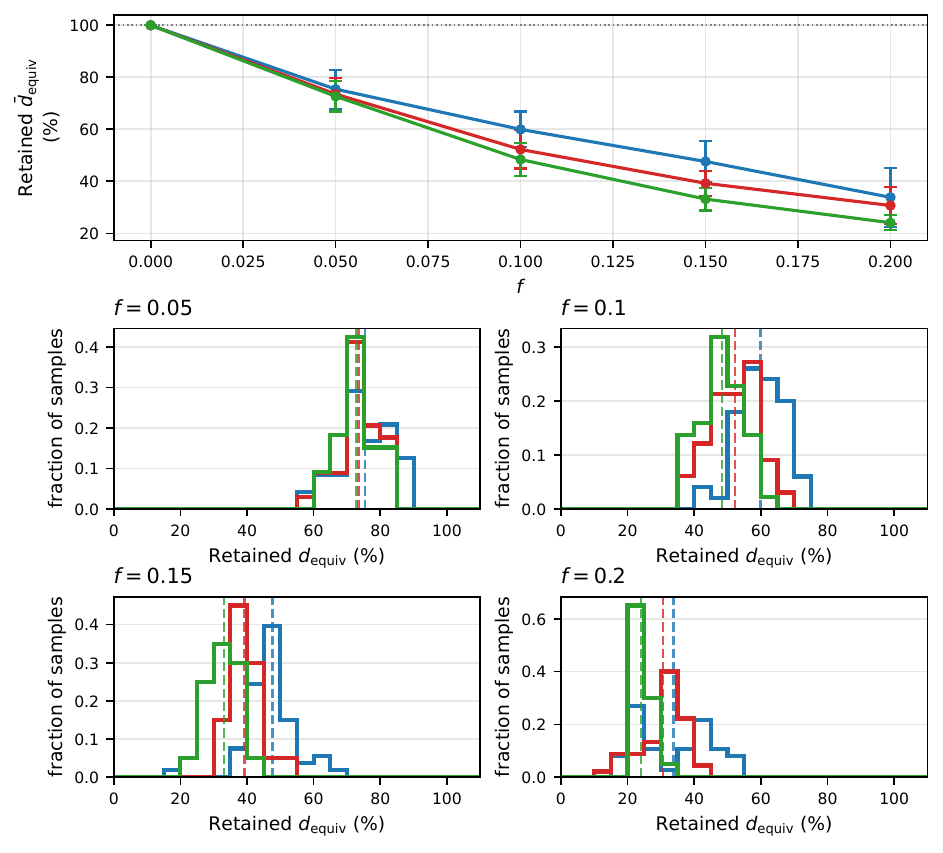}
  \caption{Equivalent distance retained under damage, as a percentage of each
  latticework's undamaged distance $d_{\max}=2G-1$, at the primary operating
  point ($p_{\rm shuttle}=2\times10^{-4}$, $p_{\rm idle}=10^{-3}$, dephasing
  shuttle channel). Blue, red and green denote the $6\times6$, $8\times8$ and
  $10\times10$ latticeworks ($d_{\max}=11,15,19$). \emph{Top:} mean retained
  $\bar d_{\rm equiv}$ vs deletion fraction $f$ ($\pm1$ std whiskers).
  \emph{Below:} per-fraction histograms of the retained $d_{\rm equiv}$, the
  three latticeworks overlaid and each normalised to its own sample count, with
  each latticework's mean drawn as a dashed vertical line.}

  \label{fig:grid-comparison}
\end{figure*}

\cref{fig:grid-comparison} and \cref{tab:retained} present the
central result. Because the equivalent distance is normalized to each grid's
undamaged distance, the three grid sizes are directly comparable.

\begin{table}[t]
  \centering
  \caption{Mean retained equivalent distance under damage, expressed as a
    percentage of each latticework's undamaged distance $2G-1$ (i.e.\ $11$,
    $15$, and $19$), at the primary operating point
    ($p_{\mathrm{shuttle}}=2\times10^{-4}$ dephasing,
    $p_{\mathrm{idle}}=10^{-3}$; X basis). Each entry averages over $20$--$53$
    damaged-latticework samples; the sample-to-sample spread and the full
    distributions are shown in \cref{fig:grid-comparison}.}
  \label{tab:retained}
  
  \begin{tabular}{lccc}
    \hline\hline
    deletion fraction $f$ & $6\times6$ & $8\times8$ & $10\times10$ \\
    \hline
    $0.05$ & $75.4\%$ & $73.5\%$ & $72.7\%$ \\
    $0.10$ & $60.0\%$ & $52.2\%$ & $48.4\%$ \\
    $0.15$ & $47.6\%$ & $39.2\%$ & $33.2\%$ \\
    $0.20$ & $33.8\%$ & $30.7\%$ & $24.1\%$ \\
    \hline\hline
  \end{tabular}
\end{table}

The overriding message is that a usable surface code survives even heavy
damage. The degradation is graceful: monotonic in the deletion fraction,
with no catastrophic-failure fraction, and a working code persists
throughout. At $5\%$ deletion, all three grids still retain roughly three
quarters of their distance. Even at $20\%$, with a fifth of the latticework
gone, every grid still hosts a functioning code --- its best distance
falling to $\approx 5$ on all three, as \cref{sec:mechanism} explains.

Across grid sizes, the larger latticeworks do retain a \emph{smaller} fraction of their distance at every deletion fraction beyond the lightest. This should not be read too pessimistically, however: the penalty grows sub-linearly in grid size. Across the $5$--$15\%$ range the incremental drop from $8\times8$ to
$10\times10$ is consistently smaller than that from $6\times6$ to $8\times8$
(e.g.\ at $10\%$, $-3.8$ points from $8\times8$ to $10\times10$ versus $-7.8$
from $6\times6$ to $8\times8$). The retained fraction thus appears to be
converging toward a grid-size-independent floor rather than continuing to fall
with each larger grid. This convergence is what makes oversizing a viable
compensation strategy: because the retained fraction levels off --- near
$\sim\!48\%$ at $10\%$ deletion --- rather than falling without bound, restoring a
target equivalent distance requires oversizing the pristine distance by a roughly
constant factor (about twofold at $10\%$ damage), independent of the target size.
Were the fraction to keep dropping with grid size, the required oversizing would
diverge and the strategy would fail. The mechanism behind both the reduction and its grid-size dependence is examined in \cref{sec:mechanism}.

\subsection{Mechanism: fitting limit and routing congestion}
\label{sec:mechanism}

Two distinct effects drive the winning distance below the undamaged maximum, and
they dominate in different damage regimes.

\begin{figure*}[t]
  \centering
  \includegraphics[width=\linewidth]{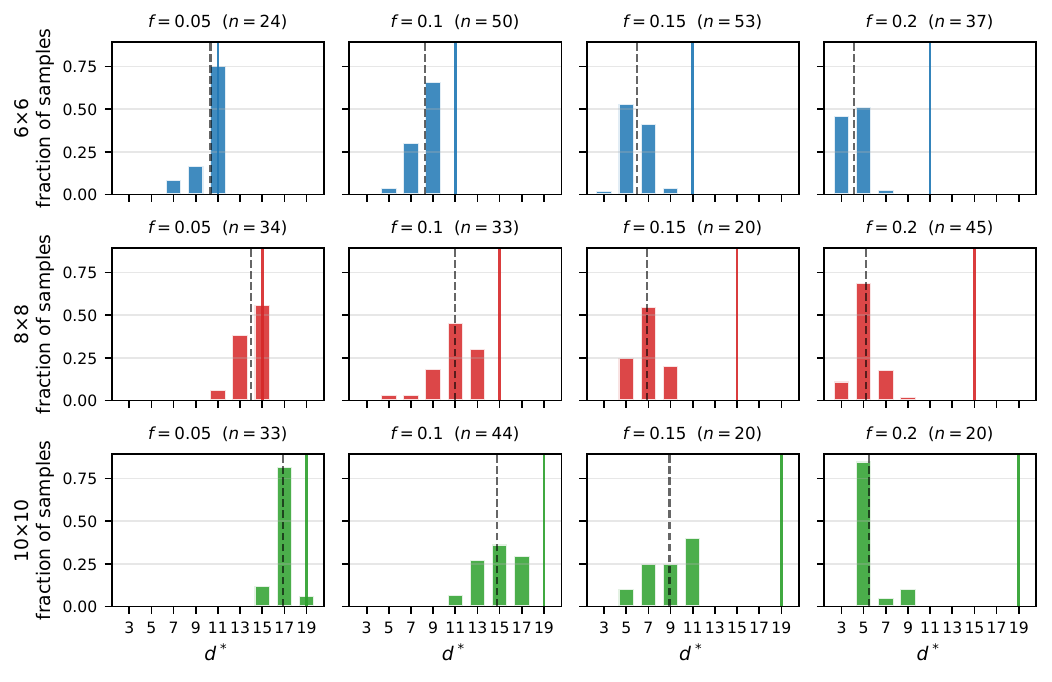}
  \caption{Distribution of the winning surface-code distance ($d^*$,
    the distance achieving the lowest LER among those that fit on the damaged
    grid) across samples. Rows are grid size, columns are deletion fraction.
    Solid line: the undamaged distance $2G-1$; dashed line: the mean
    $d^*$.}
  \label{fig:best-distance}
\end{figure*}

\begin{figure*}[t]
  \centering
  \includegraphics[width=\linewidth]{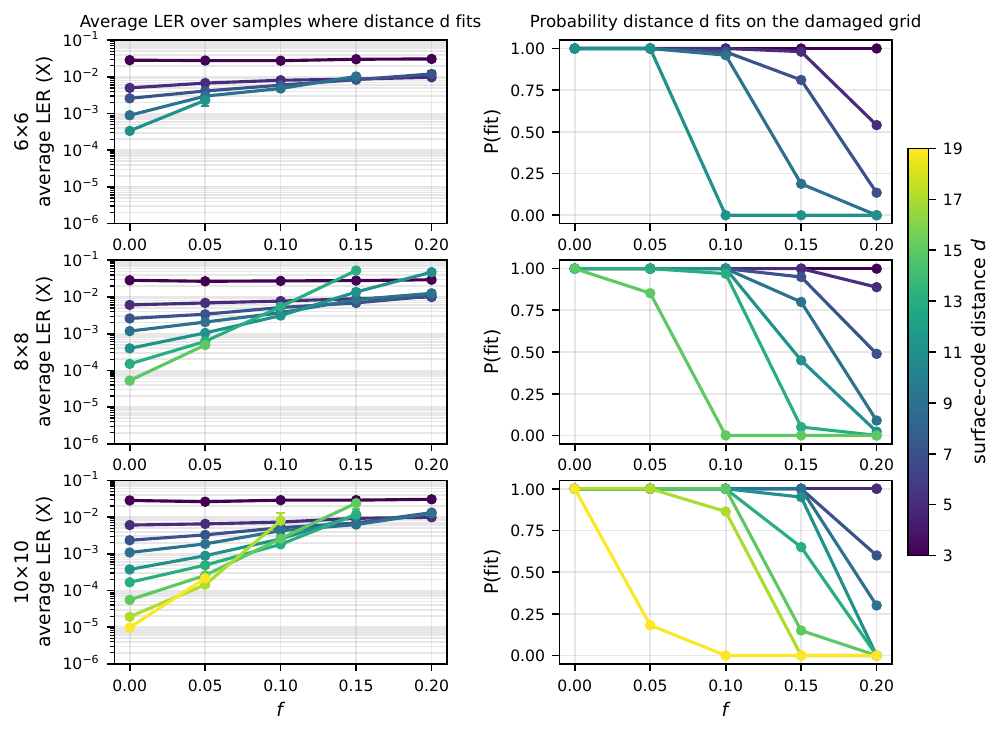}
  \caption{Per-distance behaviour vs deletion fraction, one line per distance
    $d$. \emph{Left:} average LER over the samples in which a distance-$d$ code
    fits. \emph{Right:} $P(\mathrm{fit})$, the probability that a distance-$d$
    code fits at all. Each LER curve simply terminates once its code stops
    fitting.}
  \label{fig:distance-ler-fit}
\end{figure*}

The winning distance falls steadily as damage increases: on the $10\times10$
grid the modal $d^*$ drops from $17$ at $5\%$ deletion (against an
undamaged $19$) to $15$, $11$, and finally $5$ at $20\%$
(\cref{fig:best-distance}). Across grids the mean $d^*$ is clearly
separated at light damage --- $10.3$, $14.0$, $16.9$ for the three grids at
$5\%$ --- and this separation holds through $10\%$ deletion but then collapses,
so that by $20\%$ all three grids sit at $d^*\approx5$ regardless of
their undamaged size. Two mechanisms produce this.

\paragraph{Regime 1 --- the largest codes cannot embed.}
The undamaged distance $2G-1$ requires the full latticework, and even light
damage usually destroys it: on the $10\times10$ grid a distance-$19$ code fits in
only $18\%$ of samples at $5\%$ deletion and in none by $10\%$; on the $8\times8$
grid distance $15$ fits in $85\%$ of samples at $5\%$ but never at $10\%$
(\cref{fig:distance-ler-fit}, right). The very largest codes are thus \emph{fitting-limited}: damage removes them as options outright, setting a ceiling on the embeddable distance.

\paragraph{Regime 2 --- congestion makes fitting codes underperform.}
Below that ceiling, the winning distance is driven down further than the fitting
limit alone can explain: larger codes that still fit are passed over in favour of
a smaller $d$ because they decode worse. A paired, same-sample comparison makes this apparent: restricting to the samples in which two distances both fit, the smaller code often has the lower LER. On the $8\times8$ grid at $10\%$ deletion, distance $13$ fits in $97\%$ of samples (\cref{fig:distance-ler-fit}) yet distance $11$ achieves a lower LER in $62.5\%$ of the samples where both fit, by a mean factor of $2.1$; on the $10\times10$ grid at $15\%$ deletion,
distance $11$ beats distance $13$ in $92\%$ of shared samples, by a factor of
$2.7$. Because a larger code only fits in the better-connected samples, this
same-sample comparison is if anything conservative --- the larger code is
averaged over an easier subset and still loses. At $5\%$ deletion, bigger is still better --- LER decreases monotonically with $d$. As damage grows, though, an interior optimum emerges: the best distance $d^{*}$ falls below the largest that fits and decreases with further damage. The optimum is therefore damage-induced.

\begin{figure*}[t]
  \centering
  \includegraphics[width=\linewidth]{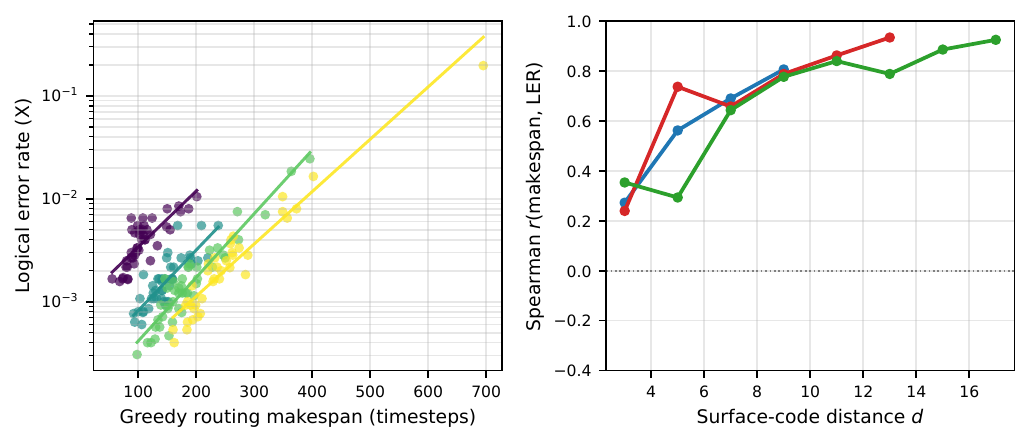}
  \caption{Routing congestion, not code size, degrades the logical error rate.
  \emph{Left:} for the $10\times10$ latticework at $f=0.1$, each damaged sample's
  LER against its greedy-schedule makespan, shown separately for four distances
  (purple $d{=}9$, teal $d{=}13$, green $d{=}15$, yellow $d{=}17$) with
  log-linear guide lines. Within every distance the LER rises with makespan, and
  the larger codes show the widest spread --- the $d{=}17$ samples range from
  $\sim\!4\times10^{-4}$ when well routed to $\sim\!2\times10^{-1}$ when badly
  routed. \emph{Right:} the makespan--LER Spearman correlation as a function of
  distance, one line per latticework (blue $6\times6$, red $8\times8$, green
  $10\times10$), still at $f=0.1$; each point is the correlation taken across the
  samples at that distance. It strengthens with code size (to $\sim\!0.9$)
  because larger codes route through more of the damaged interior.}

  \label{fig:congestion}
\end{figure*}

The proximate cause is congestion. \cref{fig:congestion}(left) takes the
$10\times10$ lattice at $f=0.1$ and, for each code distance, plots every
damaged sample's LER against its makespan, with the noise fixed at the
nominal values. At a fixed code distance, where samples differ only in which nodes were deleted, those with a longer makespan decode systematically worse. Damage forces the ancillas onto longer, more contended routes, and because idling is modelled as pure
dephasing accumulated over the schedule, the longer makespan exposes every
waiting qubit to more dephasing. Across the $f=0.1$ samples, the Spearman
correlation between makespan and LER is positive for every grid size and
distance, reaching up to $r\approx0.9$ (\cref{fig:congestion}, right). It
strengthens monotonically with code size, since a larger code routes more
ancillas through the damaged interior --- so the large codes are precisely
those whose LER is most inflated by congestion. This is why the winning
distance falls well below the largest code that geometrically fits, and why
the three grids converge on the same $d^*\approx5$ at $20\%$ deletion: the
fitting ceiling drops and, beneath it, the congestion optimum decreases until
grid size no longer confers any advantage.

\subsection{Effect of improved shuttling fidelity}
\label{sec:improved-shuttling}

\begin{figure}[t]
  \centering
  \includegraphics[width=\linewidth]{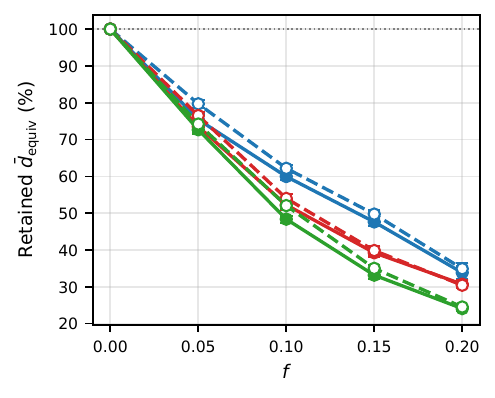}
  \caption{Retained equivalent distance $\bar d_{\rm equiv}$ (as a \% of
  $d_{\max}$) vs deletion fraction $f$. Blue, red and green denote the
  $6\times6$, $8\times8$ and $10\times10$ latticeworks; solid lines with filled
  markers are the primary shuttle rate $2\times10^{-4}$, dashed lines with open
  markers the prospective improved rate $10^{-4}$ (both dephasing). Error bars
  are the SEM.}

  \label{fig:improved-shuttling}
\end{figure}

Finally, \cref{fig:improved-shuttling} asks how much a better shuttle fidelity
would help. Halving the shuttle error from $2\times10^{-4}$ to $10^{-4}$ lifts
the retained equivalent distance by a few percentage points across the board ---
for example, at $10\%$ deletion the $10\times10$ latticework improves from
$48.4\%$ to $52.1\%$ and the $6\times6$ from $60.0\%$ to $62.2\%$ --- a modest
but genuine gain that leaves the qualitative picture of \cref{sec:main-result}
intact.

That the improvement is only modest \emph{at this operating point} is expected
rather than fundamental. We deliberately adopted a comparatively pessimistic
idling rate (\cref{sec:operating-point}), and the weak shuttle-rate sensitivity
seen here, together with the strong idling sensitivity of \cref{fig:idle-effect},
indicates that the error budget is dominated by the idle dephasing accumulated
over the schedule --- accrued by the routed ancillas and by every data qubit
waiting out the round --- rather than by the per-hop shuttle error; halving the
latter therefore moves the result only a little. This does not detract from the
results above, which were obtained at a deliberately conservative operating
point: improving the shuttle fidelity does help, and would help more where idling
is less dominant, while a better idling rate would only improve on the figures we
report.

\section{Discussion}
\label{sec:discussion}

The results identify two connectivity-rooted mechanisms behind the cost of a
damaged shuttling latticework, and both point back to the same underlying
picture. A damaged grid has a percolation threshold: once enough nodes are
deleted, the survivors no longer form a single connected component, and
restricting to the largest connected component (as we do) discards the fragments
that break away. But a functioning surface code demands more than bare
connectivity: it needs the surviving latticework to be \emph{well} connected,
with enough spare corridors to route all of its ancillas without contention.
That effective threshold is reached well before the percolation threshold. As
connectivity degrades, the largest codes first cease to embed at all
(Regime~1), and then --- for the codes that do embed --- routing grows congested, makespans lengthen, idle dephasing accumulates, and the achievable LER rises (Regime~2, \cref{fig:congestion}). The congestion effect is the more consequential of the two over most of the damage range: it is what pushes the winning distance well below the largest code that geometrically fits.

This reframes the headline trends of \cref{sec:main-result}. Larger grids
retain a smaller \emph{fraction} of their distance not because they are
inherently more fragile but because a larger grid has proportionally more of its structure in the interior, where deletions sever the corridors a large code must route through; the same deletion fraction therefore lands proportionally more damage in the costly, congestion-prone bulk. Yet the penalty grows sub-linearly in grid size and appears to converge, and in absolute terms a larger grid never does worse --- it simply stops doing \emph{better} once damage is heavy enough. Below $\sim\!10\%$ deletion, scaling the latticework buys a proportionally larger code (a $10\times10$ grid delivers $d\approx15$ where a $6\times6$ delivers $d\approx8$); by $20\%$ deletion the grids have converged on $d\approx5$ and the extra hardware earns nothing. Because the retained fraction converges rather than falling without bound, damage can nonetheless be compensated by oversizing: at
$10\%$ deletion the roughly $2\times$ oversizing implied by the $\sim\!48\%$
large-array floor is size-independent, so a designer can provision for it once rather than paying an ever-growing penalty. The practical message is therefore encouraging but bounded: the architecture degrades gracefully and keeps a working code throughout, but there is a damage ceiling beyond which a larger grid cannot be leveraged. This ceiling is set by the shuttling network's connectivity acting through the decoherence accrued over the longer
schedules damage forces; in the idealized limit of perfect, instantaneous
shuttling, only the fitting limit (Regime~1) would remain.

These results should also be read against in-place quarantine methods.
Leroux et al.~\cite{leroux2025snakes} report that, for the surface code with
$1\%$ damage to all components, the effective distance settles to $\sim\!80\%$ of
ideal, but degrades precipitously beyond $\sim\!1\%$ as valid patch solutions
become hard to find. The shuttling latticework tolerates damage of a
qualitatively different order: at $5\%$ deletion we retain $73$--$75\%$ of the
equivalent distance, and a working code persists through $10\%$ (retaining
$\sim\!48$--$60\%$) and on to $20\%$, degrading gracefully throughout. The
mobility of the ancillas is what buys this margin --- rather than walling off a
damaged region with new stabilisers on fixed hardware, the route solver simply
steers ancillas around defects, paying for continued operation in a longer, more
congested schedule (\cref{sec:mechanism}) rather than in the abrupt loss of a
legal code. We caution that the two damage models are not directly commensurate
--- Leroux et al.\ delete data qubits, ancillas and couplers at a common rate,
whereas we delete latticework nodes of all types --- so the comparison is
qualitative; but the order-of-magnitude difference in tolerable damage is not
sensitive to that distinction.

Several caveats bound these conclusions. We consider a single damage model ---
uniform random node deletion followed by restriction to the largest connected
component --- so correlated or clustered defects are not captured. Decoding is
minimum-weight perfect matching and the equivalent-distance analysis uses the
logical-$X$ observable, chosen for its direct sensitivity to the dephasing that
shuttling introduces. Most importantly, routing uses a \emph{greedy} scheduler, so our
makespans are upper bounds: an optimal scheduler would relieve some of the
congestion documented in \cref{fig:congestion}, raising the optimum
distance and recovering part of the large-code performance. The congestion penalty
we report is thus pessimistic, and better routing --- not just better qubits ---
is a concrete lever for tolerating more damage. Finally, the operating points were
fixed at the knee of the shuttle sweep and the low edge of the idling tipping
region (\cref{sec:operating-point}), so these findings hold under
deliberately conservative noise.

\subsection{Future directions}
\label{sec:future}

The present study is deliberately narrow --- one code, one damage model, one
scheduler --- and each of those choices opens a direction for further work.

\begin{itemize}
  \item \textbf{Other qLDPC codes.} Following the CAbLECAR
    programme~\cite{cablecar2026}, the same latticework and routing machinery can
    host any qLDPC code. Comparing how codes of differing rate and connectivity
    withstand damage would reveal whether the graceful degradation seen here is
    specific to the surface code or generic, and which codes best exploit --- or
    most suffer on --- a damaged latticework.
  \item \textbf{Latticework variations.} The truncated-square latticework is one
    choice among many. Different tilings, higher vertex connectivity, or
    additional emitter/receiver nodes would change both the percolation threshold
    and the routing slack available to absorb damage, and may shift the
    congestion-limited regime of \cref{sec:mechanism}.
  \item \textbf{Constrained directional control.} We allow each edge to carry an
    ancilla either way, scheduled per path. Fixing a single permitted direction
    per edge --- likely easier to realise experimentally --- lowers the control
    overhead at the cost of routing freedom; quantifying that trade-off under
    damage is a natural next step.
  \item \textbf{Asynchronous syndrome extraction.} The makespan is often set by
    one or a few slow stabilisers. Rather than stalling the whole schedule,
    stabilisers could be measured as soon as they are ready, so that slow
    stabilisers are simply measured less often without holding up the rest.
    Since the damage penalty is paid chiefly in dephasing accumulated over the
    lengthened schedule (\cref{sec:mechanism}), decoupling the round time
    from the slowest route attacks exactly the makespan-times-dephasing product
    that congestion inflates. More broadly, alternative syndrome-extraction
    circuits could be explored.
  \item \textbf{Penalise rather than delete.} Instead of removing damaged nodes
    outright, they could be down-weighted: an ancilla might be forbidden to
    \emph{wait} at a damaged node but allowed to pass through it, or allowed to
    pass only at reduced speed. This softer damage model may recover much of the
    distance that hard deletion discards.
  \item \textbf{Non-uniform damage.} Real defects may be spatially correlated or
    clustered rather than uniformly random. Clustered damage could be either more
    forgiving (leaving large clean regions) or more harmful (severing whole
    corridors), and is worth studying directly.
  \item \textbf{Continuous-time model.} Our discrete model advances every ancilla
    by at most one hop per timestep. A continuous-time treatment, closer to
    CAbLECAR~\cite{cablecar2026}, would capture heterogeneous hop durations and
    finer-grained scheduling, and could sharpen the congestion analysis.
  \item \textbf{Better solvers.} Our greedy scheduler yields only an upper bound on
    the makespan. Optimal or near-optimal routing would give shorter, less
    contended schedules, letting the larger codes keep their advantage further into
    the damage range --- raising the best distance and hence the tolerable
    damage.
  \item \textbf{Circuit tailoring.} As in CAbLECAR~\cite{cablecar2026},
    conjugating the CNOTs on the $Z$-check ancillas with Hadamards would let those
    ancillas traverse their routes in a basis insensitive to the shuttle
    dephasing, greatly improving the effective shuttling fidelity at essentially
    no routing cost --- a nearly free gain in our error model.
  \item \textbf{Logical operations.} We have studied memory only. How
    logical operations --- lattice surgery or code deformation --- fare
    on a damaged latticework is an important open question for a full
    architecture (logical-level, defect-aware routing on shuttling lattices is beginning to be explored~\cite{shen2026routing}).
      \item \textbf{Runtime emergent damage.} We have assumed that damage is entirely static, and as such all analysis is offline. The challenge of updating routing solutions mid execution of a quantum algorithm is both important and very non-trivial.

\end{itemize}

\begin{figure*}
  \centering
  \includegraphics[width=\linewidth]{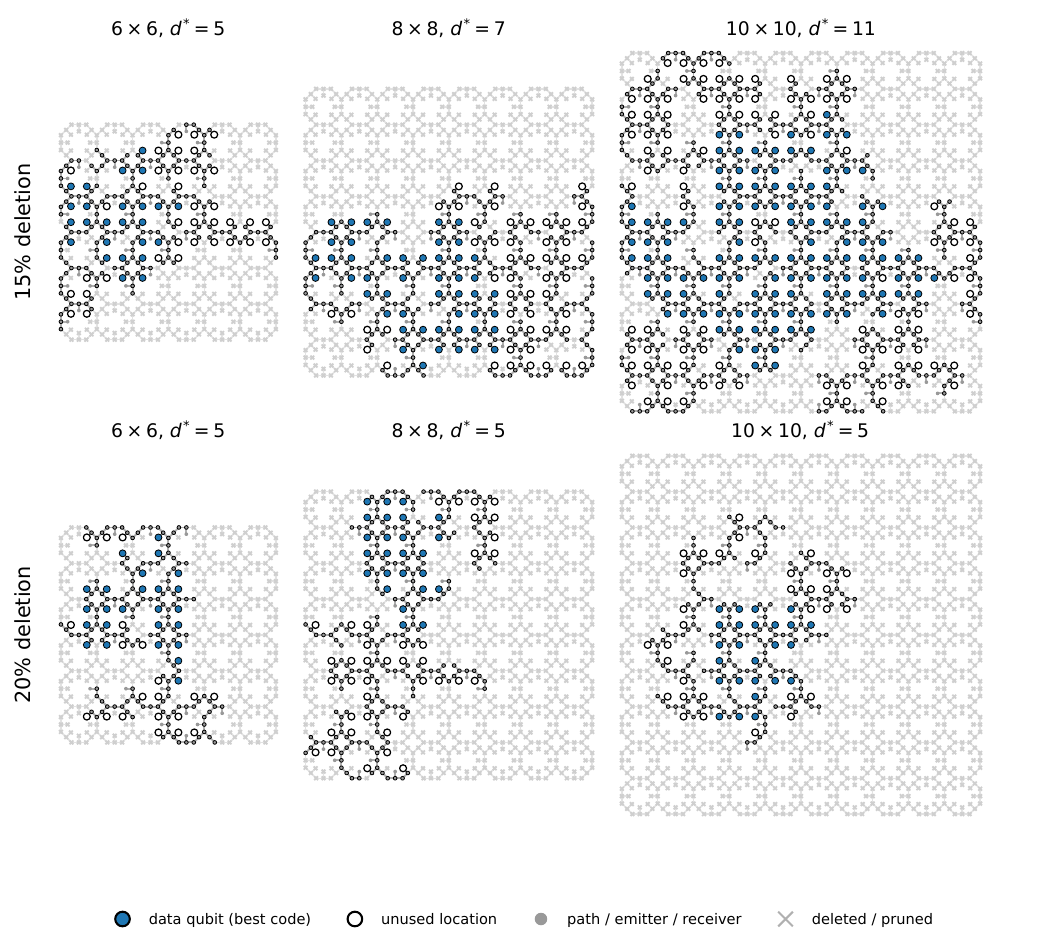}
  \caption{Best-code placement across the fragmentation threshold; all six panels
  share a common cell size. Each shows one damaged instance --- the data-qubit
  sites of the lowest-LER distance $d^{*}$ filled blue on the surviving
  latticework, other surviving locations white (unused), emitter/receiver/path
  nodes as faint grey dots, and deleted or pruned material as faint crosses.
  \emph{Top ($15\%$ deletion):} below the fragmentation threshold the surviving
  network is only lightly broken up, and the best code grows with grid size
  ($d^{*}=5,7,11$ for $6\times6$, $8\times8$, $10\times10$) --- a larger array
  hosts a larger code. \emph{Bottom ($20\%$ deletion):} as fragmentation becomes dominant the network breaks into well-connected islands possibly joined by thin corridors;
  even when a larger code would still fit, it is not selected, since spanning
  several islands would route through the congested bottlenecks. The best code is
  then confined to a single island ($d^{*}=5$ on all three grids here), and the
  grids converge (\cref{fig:best-distance}).}
\label{fig:best-placement}
\end{figure*}

\section{Conclusion}
\label{sec:conclusion}

We have characterised the rotated surface code on damaged shuttling
latticeworks, using a circuit-level model with conventional gate noise and a
physically motivated pure-dephasing description of shuttling and idling, and
evaluated it at deliberately conservative operating points
($p_{\mathrm{shuttle}}=2\times10^{-4}$, $p_{\mathrm{idle}}=10^{-3}$). The
central finding is an encouraging one: the architecture tolerates damage well
beyond the level at which fixed-hardware quarantine schemes lose viability.
Across $6\times6$, $8\times8$ and $10\times10$ latticeworks the code degrades
gracefully and continuously --- with no catastrophic-failure fraction --- and a
working logical qubit survives across the entire range we studied, from a few
percent deletion up to $20\%$, where every grid still manages to a host a meaningful
code. At $10\%$ damage the equivalent distance is reduced only to
$\sim\!48$--$60\%$ of its pristine value; larger arrays lose more of their potential, but because this fraction converges as the
array grows rather than falling without bound, the loss can be absorbed by
oversizing the array by a roughly constant factor --- about twofold linear dimension at $10\%$.
In-place quarantine methods, by contrast, lose viability already beyond
$\sim\!1\%$ damage, so the mobility of the ancillas extends the tolerable damage
by an order of magnitude.

This margin may be crucial for near-term hardware, where it may make functional QEC possible on devices that would otherwise be too immature for use. The
first large spin-qubit arrays will not be fabricated at the near-unity component
yields that mature fault tolerance assumes; defect and dropout rates of order a
few to ten percent may be a phase that early devices pass through even when lower damage rates are ultimately possible. A shuttling latticework therefore offers a concrete route to utilising
early, imperfectly-yielded processors --- at a hardware overhead
(oversizing) that is bounded and predictable rather than divergent --- rather
than demanding the sub-percent defect rates required to stay within the reach of
in-place quarantine.

The degradation in logical performance versus damage fraction is gradual and can
be intuitively understood through a few key observations. It is obvious that, as
damage increases, one loses the capability to host a surface code of a given
size, simply because there are no longer sufficient qubit host locations;
however, a network that can \emph{barely} host a code of a given distance will
sometimes achieve a better LER with a smaller code, because of the congestion
incurred by the larger one (see e.g.\ the left panels of
\cref{fig:distance-ler-fit}). At high enough damage this congestion acquires a
clear geometric origin: the surviving latticework, though still a single
connected component, effectively fragments into well-connected islands linked to
one another by only a few surviving corridors --- well before it would truly
disconnect. A code large enough to span several islands must route through these
bottlenecks, so the best usable distance comes to be set by the size of a single
island rather than by the total lattice. This is seen directly in where the best code is placed (\cref{fig:best-placement}): below the threshold (e.g.\ $15\%$ deletion) the best
code still grows with grid size, whereas at $20\%$ it is confined to a single
well-connected island of $\sim25$ sites ($d^{*}=5$) on all three grids, the rest
of the surviving network left unused. Because that size is a local,
damage-determined scale, it is largely independent of the overall array ---
consistent with all three grids converging on the same $d^{*}\approx5$ at $20\%$
damage, which we read as an early indication of this regime. We call the damage
level at which it sets in the \emph{fragmentation threshold}.

Our results indicate that at damage levels below this fragmentation threshold, one can compensate for damage by oversizing the original array. Various factors can reduce the cost, to differing extents. At our deliberately pessimistic idling operating
point, improving the per-hop shuttle fidelity alone recovers comparatively
little --- the binding cost is the schedule rather than the individual hop ---
though a better idling rate would itself raise performance. Better scheduling in
particular, since our greedy router only upper-bounds the makespan, is a direct
and as-yet unused lever for pushing the tolerable damage higher still. The
approach also generalises readily: the same latticework and routing machinery
host any qLDPC code, so per-code damage robustness is an immediate extension, and
while we have treated damage as static and known before runtime, the
pre-computation of schedules could be repeated to adapt to defects that emerge
during operation. We leave these, along with the other
variations listed in \cref{sec:future}, to future work.

\begin{acknowledgments}

Q.E. acknowledges the support of the Research Foundation-Flanders through the Fundamental Research PhD programme (grant no. 11Q4A24N), as well as the EOS-FWO-FNRS project CHEQS.
\end{acknowledgments}

\section*{Author Contributions}
S.C.B proposed the study, produced early versions of the routing code and co-wrote the paper. Q.E. produced all the final software tools, created the Stim-based QEC simulations, performed the numerical studies, and defined and answered the detailed questions reported in the paper.

\section*{Competing interests}
The authors declare no competing interests.

\vspace{1cm}

\bibliography{ref}

\appendix
\makeatletter 
\renewcommand\thefigure{\thesection.\arabic{figure}} 
\renewcommand{\theHfigure}{\thesection.\arabic{figure}}
\makeatother



\end{document}